\documentclass[10pt,a4paper,prb,twocolumn,nofootinbib]{revtex4}
\usepackage[latin9]{inputenc}
\usepackage{amsmath}
\usepackage{amssymb}
\usepackage{graphicx}

\makeatletter

\pdfpageheight\paperheight
\pdfpagewidth\paperwidth

\@ifundefined{textcolor}{}
{%
 \definecolor{BLACK}{gray}{0}
 \definecolor{WHITE}{gray}{1}
 \definecolor{RED}{rgb}{1,0,0}
 \definecolor{GREEN}{rgb}{0,1,0}
 \definecolor{BLUE}{rgb}{0,0,1}
 \definecolor{CYAN}{cmyk}{1,0,0,0}
 \definecolor{MAGENTA}{cmyk}{0,1,0,0}
 \definecolor{YELLOW}{cmyk}{0,0,1,0}
}

\pdfoutput=1

\usepackage{physics}
\usepackage{dcolumn}
\usepackage{bm}

\makeatother

\begin{document}

\title{Nonlinear optical responses of crystalline systems: \\
 Results from a velocity gauge analysis}

\author{D. J. Passos}
\email{passos.djs@gmail.com}

\affiliation{Centro de Física das Universidades do Minho e Porto}

\affiliation{Departamento de Física e Astronomia, Faculdade de Ciências, Universidade
do Porto, 4169-007 Porto, Portugal}

\author{G. B. Ventura}

\affiliation{Centro de Física das Universidades do Minho e Porto}

\affiliation{Departamento de Física e Astronomia, Faculdade de Ciências, Universidade
do Porto, 4169-007 Porto, Portugal}

\author{J. M. Viana Parente Lopes}

\affiliation{Centro de Física das Universidades do Minho e Porto}

\affiliation{Departamento de Física e Astronomia, Faculdade de Ciências, Universidade
do Porto, 4169-007 Porto, Portugal}

\author{J. M. B. Lopes dos Santos}

\affiliation{Centro de Física das Universidades do Minho e Porto}

\affiliation{Departamento de Física e Astronomia, Faculdade de Ciências, Universidade
do Porto, 4169-007 Porto, Portugal}

\author{N. M. R. Peres}

\affiliation{Centro de Física das Universidades do Minho e Porto}

\affiliation{Departamento de Física, Universidade do Minho, P-4710-057, Braga,
Portugal}
\begin{abstract}
In this work, the difficulties inherent to perturbative calculations
in the velocity gauge are addressed. In particular, it is shown how
calculations of nonlinear optical responses in the independent particle
approximation can be done to any order and for any finite band model.
The procedure and advantages of the velocity gauge in such calculations
are described. The addition of a phenomenological relaxation parameter
is also discussed. As an illustration, the nonlinear optical response
of monolayer graphene is numerically calculated using the velocity
gauge. 
\end{abstract}
\maketitle
\global\long\def\ket#1{\left|#1\right\rangle }
 \global\long\def\bra#1{\left\langle #1\right|}

\global\long\def\braket#1#2{\langle#1|#2\rangle}

\section{INTRODUCTION}

A common approach to understanding nonlinear optical effects in atomic
and condensed matter systems comes from a perturbation theory where
the electric current is expanded in powers of an external applied
electric field \cite{shen1984principles}, assumed to be sufficiently
weak for the expansion to be physically meaningful. In this framework,
one attempts to calculate nonlinear optical response functions, usually
second or third order susceptibilities, the different frequency components
of which describe a variety of physical phenomena like the Kerr effect,
harmonic generation, electro-optic effect, \emph{etc.}

The theory was first developed for atomic systems and, in the early
nineties \cite{sipe1993nonlinear,aversa1995nonlinear}, extended to
crystalline systems, characterized by electronic bands and the corresponding
Bloch functions. The simpler calculations follow two essential assumptions
which will be adopted throughout this work: the independent particle
approximation, where explicit electron-electron interactions are disregarded
(except possibly in the determination of the electron bands, as in
Density Functional Theory), and the electric dipole approximation,
where the response functions are taken to be local in space, a consequence
of the long wavelength limit of the applied electric field.

Even in this simple approach, however, difficulties were found regarding
different ways to describe the perturbation. We can  write the complete
Hamiltonian of the crystal under the influence of the external field
in two ways, either in the length gauge,

\begin{equation}
H(t)=H_{0}(\textbf{r},\textbf{p})-q\,\textbf{r}\cdot\textbf{E}(t),
\end{equation}
or in the velocity gauge or minimal coupling Hamiltonian,

\begin{equation}
H(t)=H_{0}(\textbf{r},\textbf{p}-q\textbf{A}(t)),
\end{equation}
where $H_{0}$ is the unperturbed crystal Hamiltonian,  $q=-e$ is
the charge of the electron and $\textbf{A}(t)$ the vector potential,
chosen so that $\textbf{E}(t)=-\partial_{t}\textbf{A}(t)$. These
two descriptions can easily be shown to be equivalent and related
by a time-dependent unitary transformation \cite{ventura2017gauge}.

The first attempts at computing nonlinear optical responses in crystals
made use of the minimal coupling method \cite{genkin1968contribution,sipe1993nonlinear},
since it has the advantage of retaining the translation symmetry of
the Hamiltonian, only coupling states with the same Bloch vector $\textbf{k}$.
However, it presented some serious difficulties as it seemed that
response functions computed in the velocity gauge diverged in the
limit of low frequencies, even for the case of zero temperature insulators,
where such divergences should clearly be absent. A more detailed analysis
of the linear response showed that these were only apparent divergences
that could be removed by careful manipulations and sum rules \cite{sipe1993nonlinear}.
As presented, these procedures were cumbersome and not easily generalizable
to higher order response functions.

This led to a preference for the length gauge in nonlinear optical
response calculations \cite{aversa1995nonlinear,hughes1996calculation,sipe2000second,al2014high,hipolito2016nonlinear}.
This gauge presents its own difficulties, the most notorious one being
that the matrix elements of the position operator in the Bloch basis
are ill-defined until the thermodynamic limit is taken and, even then,
they can be defined only as a distribution. Inspired by Blount's work
\cite{blount1962formalisms}, Aversa et al. \cite{aversa1995nonlinear}
pointed out that the position operator will appear in the calculations
only inside commutators whose matrix elements are well defined and
successfully developed the theory in the length gauge. This formalism
has since been applied to various systems \cite{cheng2014dc,cheng2015third,hipolito2016nonlinear,hughes1996calculation,nastos2005scissors,al2014high}.

More recently, and still within the length gauge approach, simple
expressions for the nonlinear conductivity to arbitrary order in the
electric field were derived by the present authors, making use of
the covariant derivative notation \cite{ventura2017gauge}. These
expressions are useful for inspection and reduced the calculation
procedure to a fairly straightforward expansion of commutators. Although
the idea of a ``generalized derivative'' was already around in the
literature \cite{aversa1995nonlinear}, it involved separating the
intra and inter-band components of the position operator and response
functions. The emphasis on these distinctions, originally motivated
by the intent of applying the equations to the special case of clean,
cold semiconductors and to make analogies with atomic and free electron
systems, made the calculations less transparent, in our view.

In this same work \cite{ventura2017gauge}, similar and equivalent
expressions were derived using the velocity gauge, for the Hamiltonian

\begin{equation}
H_{0}(\mathbf{r},\mathbf{p})=\frac{\mathbf{p}^{2}}{2m}+V_{L}(\mathbf{r}),\label{free}
\end{equation}
where the second term is the periodic lattice potential. It was shown
that, in the form derived from the velocity gauge, the expressions
for the nonlinear optical coefficients lose their validity if only
a finite number of bands around the Fermi level are taken into account.
This difficulty was recognized early on \cite{aversa1994third,aversa1995nonlinear},
and, together with the apparent infrared divergences, led to the velocity
gauge being less adopted. The reasons for the failure of these expressions
under a truncation in band space were also understood \cite{rzkazewski2004equivalence,aversa1995nonlinear}
and recently subjected to a more quantitative analysis \cite{taghizadeh2017linear}.
One of the arguments \cite{aversa1995nonlinear} consisted in noting
that the sum rules that connected the expressions in the two gauges
seem to rely on commutator identities such as

\begin{equation}
\left[r^{\alpha},v^{\beta}\right]=\frac{i\hbar}{m}\,\delta^{\alpha\beta}\label{identity}
\end{equation}
($\mathbf{r}$ and $\mathbf{v}$, position and velocity operators),
which require an infinite number of bands to hold. This led to a common
misconception that the velocity gauge could only be properly implemented
if an infinite number of bands is taken into account \cite{aversa1994third,aversa1995nonlinear,virk2007semiconductor,cheng2014third,al2014high,taghizadeh2017linear}.
In fact, sum rules of greater generality have been constructed which
remain true even under truncation of the bands \cite{ventura2017gauge}.
Nonetheless, various authors have indicated that the velocity gauge,
if employed, would lead to different, unphysical, predictions \cite{virk2007semiconductor,taghizadeh2017linear}.

The fundamental difference between the two gauges that seems not to
have been properly appreciated concerns the form of the perturbation.
In the velocity gauge, \textit{the form of the perturbation depends
explicitly on} $H_{0}$, unlike in the case of the length gauge. Expressions
for the nonlinear coefficients derived from the Hamiltonian in Eq.~\ref{free},
cannot be truncated to a finite set of bands. A truncation of $H_{0}$
implies a different form of the perturbation, and leads to different
expressions for the nonlinear conductivities in the velocity gauge,\emph{
but not} in the length gauge.

In order to build the theory in its most general form, we will make
no assumptions on the form of $H_{0}$, other than it has the periodicity
of some Bravais lattice, so that Bloch's theorem applies and there
is a well defined First Brillouin Zone (FBZ). The derived forms for
the nonlinear conductivities will be completely equivalent to the
ones obtained from the length gauge and can be applied to any finite
band model; no commutator identities of the kind of Eq.~\ref{identity}
are assumed.

In the next section, the equivalence of the two gauges is revisited,
starting from an Hamiltonian with a finite set of bands, and some
formal concepts behind the velocity gauge formulation are reviewed.
In section~\ref{sec:RECURSIVE-SOLUTIONS-TO}, the equation of motion
for the reduced density matrix in the velocity gauge is introduced
in a form more general than previously presented. Recursively solving
the equation of motion leads to the nonlinear conductivities. The
advantages and subtleties of using this gauge are discussed in Section~\ref{sec:LENGTH-VS-VELOCITY},
where it is argued that the velocity gauge should prove more efficient
for numerical computations. In Section~\ref{sec:PHENOMENOLOGICAL-RELAXATION-PARA},
the introduction of a phenomenological relaxation parameter is addressed,
and proves to be less trivial than expected. As an example, in Section~\ref{sec:NUMERICAL-IMPLEMENTATION}
the formalism is applied to third harmonic generation in graphene,
in a full Brillouin zone calculation, and a comparison is made with
already existing results in the literature. Section~\ref{sec:CONCLUSIONS}
contains some closing remarks.

\section{DENSITY MATRIX FORMALISM\label{sec:DENSITY-MATRIX-FORMALISM}}

In very general terms, which do not exclude finite band models, the
single particle unperturbed Hamiltonian $H_{0}$ can be written as

\begin{equation}
H_{0}=\sum_{s}\int\!\frac{d^{d}\textbf{k}}{(2\pi)^{d}}\ket{\psi_{\textbf{k}s}}\left[H_{0}\right]_{\mathbf{k}ss}\bra{\psi_{\textbf{k}s}},\label{H0}
\end{equation}
where $\ket{\psi_{\textbf{k}s}}$ are the Bloch band states and $\left[H_{0}\right]_{\mathbf{k}ss'}\equiv\epsilon_{\mathbf{k}s}\delta_{ss'}$,
with $\epsilon_{\mathbf{k}s}$ the band energies. 

To represent the scalar potential, $-q\,\mathbf{E}\cdot\mathbf{r}$,
in the Bloch basis, we require from the start the infinite volume
limit. We define the normalization of the Bloch wave functions as

\begin{equation}
\braket{\psi_{\mathbf{k}'s'}}{\psi_{\mathbf{k}s}}=\left(2\pi\right)^{d}\delta_{ss'}\delta\left(\mathbf{k}-\mathbf{k}'\right).
\end{equation}
where $d$ is the dimensionality of the system. Using Blount's results
for the matrix elements of the position operator \cite{blount1962formalisms},

\begin{align}
\bra{\psi_{\mathbf{k}s}}\mathbf{r}\ket{\psi_{\mathbf{k's'}}} & =(2\pi)^{d}\left[\delta_{ss'}(-i)\nabla_{\mathbf{k}'}\delta(\mathbf{k}'-\mathbf{k})\right.\nonumber \\
 & +\left.\delta(\mathbf{k}-\mathbf{k}')\,\boldsymbol{\xi}_{\mathbf{k}ss'}\right],
\end{align}
where $\xi_{\textbf{k}ss'}$ is the Berry connection \cite{berry1984quantal,xiao2010berry},
one obtains for the single particle perturbed Hamiltonian in the length
gauge \cite{ventura2017gauge}(see Appendix \ref{B})

\begin{equation}
H^{E}=\int\frac{d^{d}\mathbf{k}}{(2\pi)^{d}}\sum_{s,s'}\ket{\psi_{\mathbf{k}s}}\left[\delta_{ss'}\epsilon_{\mathbf{k}s}-iq\mathbf{E}(t)\cdot\mathbf{D}_{\mathbf{k}ss'}\right]\bra{\psi_{\mathbf{k}s'}}\label{eq:He}
\end{equation}
where the covariant derivative, $\mathbf{D}_{\mathbf{k}ss'},$ is
defined by

\begin{equation}
\mathbf{D}_{\mathbf{k}ss'}\equiv\delta_{ss'}\nabla_{\mathbf{k}}-i\boldsymbol{\xi}_{\mathbf{k}ss'}.\label{eq:covDeriv}
\end{equation}

In our previous paper \cite{ventura2017gauge}, we discussed the equivalence
of the length and velocity gauges, starting from a theory with an
infinite number of bands (Eq.~\ref{free}); we showed that a suitable
truncation of final expressions for the nonlinear optical response
functions to a finite set of bands leads to a reasonable approximation
\emph{only }in the length gauge. Therefore, if we want to formulate
correctly a velocity gauge calculation with a finite set of bands,
we should take Eqs.~\ref{eq:He} and \ref{eq:covDeriv} as our starting
point, and obtain the single particle velocity gauge Hamiltonian,
$H^{A}$, from a time dependent unitary transformation of $H^{E}$.
In this fashion, we preserve the equivalence of both descriptions,
even when the sum over band indexes is finite. In appendix \ref{B}
we derive

\begin{equation}
H^{A}=\sum_{s,s'}\int\frac{d^{d}k}{\left(2\pi\right)^{d}}\ket{\psi_{\mathbf{k}s}}H_{\mathbf{k}ss'}^{A}\bra{\psi_{\mathbf{k}s'}}
\end{equation}
with

\begin{equation}
H_{\mathbf{k}ss'}^{A}=\epsilon_{\textbf{k}s}\,\delta_{ss'}+V_{\textbf{k}ss'}(t),
\end{equation}
the time dependent perturbation, $V_{\textbf{k}ss'}(t)$, being expressed
as a power series in the vector potential, 
\begin{equation}
V_{\mathbf{k}ss'}(t)=\sum_{n=1}^{\infty}\frac{(-q)^{n}\,A_{\alpha_{1}}(t)\dots A_{\alpha_{n}}(t)}{n!}\,h_{\mathbf{k}ss'}^{\alpha_{1}\dots\alpha_{n}},\label{expansion-1}
\end{equation}
where

\begin{equation}
h_{\mathbf{k}ss'}^{\alpha_{1}\dots\alpha_{n}}\equiv\hbar^{-n}\left[D_{\mathbf{k}}^{\alpha_{n}},\left[\dots,\left[D_{\mathbf{k}}^{\alpha_{1}},H_{0}\right]\right]...\right]_{ss'}.\label{h-1}
\end{equation}
An implicit summation over repeated Cartesian components $\alpha_{i}$
is henceforth assumed. The coefficients in the expansion are written
explicitly in Eq.~\ref{h-1} in terms of commutators involving covariant
derivatives.

If we take the coefficient of the first order term in the expansion
of Eq.~\ref{h-1} as an example,

\begin{equation}
h_{\textbf{k}ss'}^{\alpha}=\hbar^{-1}\,\left[D_{\mathbf{k}}^{\alpha},H_{0}\right]_{ss'}=\frac{1}{\hbar}\frac{\partial\epsilon_{\mathbf{k}s}}{\partial k_{\alpha}}\,\delta_{ss'}-\frac{i}{\hbar}\,\xi_{\mathbf{k}ss'}^{\alpha}(\epsilon_{\mathbf{k}s'}-\epsilon_{\mathbf{k}s})
\end{equation}
These are the unperturbed velocity matrix elements, since

\begin{equation}
\textbf{v}_{\textbf{k}ss'}^{(0)}=-i\hbar^{-1}\left[\textbf{r},H_{0}\right]_{\textbf{k}ss'}=\hbar^{-1}\left[\mathbf{D}\textbf{\ensuremath{_{\mathbf{k}}}},H_{0}\right]_{ss'}=\textbf{h}_{\textbf{k}ss'}.\label{eq:velocity1}
\end{equation}

Had we started from the Hamiltonian of Eq.~\ref{free}, the perturbation
expansion of Eq.~\ref{expansion-1} would be reduced to the linear
term in $\mathbf{A}(t)$; the second order term would have been a
$\mathbf{k}$ independent constant, irrelevant for the dynamics, and
all higher order terms would have been zero \cite{ventura2017gauge}.
But to proceed in more general terms and ensure equivalence between
length and velocity gauges for finite band models, we must, for now,
retain all the terms in the series.

For the analysis of the steady state response of the electric current
density, \textbf{J}, it is useful to rewrite the perturbation of Eq.~\ref{expansion-1}
in frequency space, where the connection can already be made with
the electric field components by $\textbf{E}(\omega)=i\omega\textbf{A}(\omega)$,

\begin{align}
V_{\textbf{k}ss'}(\omega)= & \sum_{n=1}^{\infty}\int_{-\infty}^{+\infty}\frac{d\omega_{1}}{2\pi}...\frac{d\omega_{n}}{2\pi}\nonumber \\
\times & \frac{(iq)^{n}\,E_{\alpha_{1}}(\omega_{1})\dots E_{\alpha_{n}}(\omega_{n})}{n!\;\omega_{1}\dots\omega_{n}}\,h_{\textbf{k}ss'}^{\alpha_{1}\dots\alpha_{n}}\nonumber \\
\times & (2\pi)\;\delta(\omega-\omega_{1}-\dots-\omega_{n})\label{eq:expansion-2}
\end{align}

Having carefully defined the perturbation in the velocity gauge, we
can turn to the nonlinear optical response functions, defined in frequency
space by a similar expansion,

\begin{align}
\langle J_{\alpha}\rangle(\omega) & =\int\frac{d\omega_{1}}{2\pi}\;\sigma_{\alpha\beta}^{(1)}(\omega_{1})E^{\beta}(\omega_{1})\,(2\pi)\,\delta(\omega-\omega_{1})\nonumber \\
 & +\int\frac{d\omega_{1}}{2\pi}\frac{d\omega_{2}}{2\pi}\;\sigma_{\alpha\beta\gamma}^{(2)}(\omega_{1},\omega_{2})E^{\beta}(\omega_{1})E^{\gamma}(\omega_{2})\nonumber \\
 & \times(2\pi)\,\delta(\omega-\omega_{1}-\omega_{2})\nonumber \\
 & +\dots\label{eq:defresponsefunction}
\end{align}
The ensemble average of the electric current density, as for any observable,
is given in terms of the density matrix $\rho$,

\begin{align}
\langle\,\textbf{J}\,\rangle(t)=\text{Tr}[\textbf{J}\,\rho(t)] & =q\int\frac{d^{d}\textbf{k}}{(2\pi)^{d}}\sum_{ss'}\,\textbf{v}_{\textbf{k}ss'}\,\text{Tr}\left[c_{\textbf{k}s}^{\dagger}c_{\textbf{k}s'}\rho(t)\right]\nonumber \\
 & =q\int\frac{d^{d}\textbf{k}}{(2\pi)^{d}}\sum_{ss'}\textbf{v}_{\textbf{k}ss'}\rho_{\textbf{k}s's}(t)
\end{align}
where the reduced density matrix (RDM) is defined as a the expectation
of a product of a creation and a destruction operator in Bloch states:

\begin{equation}
\rho_{\textbf{k}s's}(t)\equiv\text{Tr}\left[c_{\textbf{k}s}^{\dagger}c_{\textbf{k}s'}\rho(t)\right]=\langle c_{\textbf{k}s}^{\dagger}c_{\textbf{k}s'}\rangle.\label{RDM}
\end{equation}
 The standard density matrix formalism computes the nonlinear conductivities
by expanding the RDM in powers of the electric field. In the velocity
gauge, however, the \textit{electric current is an explicitly time
and field dependent observable}. The velocity matrix elements are
defined by $-i\hbar^{-1}\left[\mathbf{r},H^{A}\right]=\hbar^{-1}\left[\mathbf{D},H^{A}\right]$
and also have to be expanded in powers of the electric field:

\begin{align}
v_{\mathbf{k}ss'}^{\beta}(\omega)= & \sum_{n=0}^{\infty}\int_{-\infty}^{+\infty}\frac{d\omega_{1}}{2\pi}...\frac{d\omega_{n}}{2\pi}\nonumber \\
\times & \frac{(iq)^{n}\,E_{\alpha_{1}}(\omega_{1})\dots E_{\alpha_{n}}(\omega_{n})}{n!\;\omega_{1}\dots\omega_{n}}\,h_{\textbf{k}ss'}^{\beta\alpha_{1}\dots\alpha_{n}}\nonumber \\
\times & (2\pi)\;\delta(\omega-\omega_{1}-\dots-\omega_{n})\label{Ref:vexpansion}
\end{align}
The expansion must therefore be done in the density matrix and the
velocity matrix elements simultaneously. In the absence of an external
field, the current is

\begin{equation}
\langle\textbf{J}\rangle^{(0)}=q\int\frac{d^{d}\textbf{k}}{(2\pi)^{d}}\sum_{ss'}\textbf{v}_{\textbf{k}ss'}^{(0)}\rho_{\textbf{k}s's}^{(0)}.\label{J0}
\end{equation}
The first order response is

\begin{equation}
\langle\textbf{J}\rangle^{(1)}=q\int\frac{d^{d}\textbf{k}}{(2\pi)^{d}}\sum_{ss'}\left(\textbf{v}_{\textbf{k}ss'}^{(1)}\rho_{\textbf{k}s's}^{(0)}+\textbf{v}_{\textbf{k}ss'}^{(0)}\rho_{\textbf{k}s's}^{(1)}\right),\label{J1}
\end{equation}
and, in general,

\begin{equation}
\langle\textbf{J}\rangle^{(n)}=q\int\frac{d^{d}\textbf{k}}{(2\pi)^{d}}\sum_{p=0}^{n}\sum_{ss'}\textbf{v}_{\textbf{k}ss'}^{(n-p)}\rho_{\textbf{k}s's}^{(p)}\label{Jn}
\end{equation}

The expansion of the velocity matrix elements in the external field
is already defined in Eq.~\ref{Ref:vexpansion}. The expansion of
the reduced density matrix involves solving its equation of motion
recursively.

\section{RECURSIVE SOLUTIONS TO THE EQUATION OF MOTION\label{sec:RECURSIVE-SOLUTIONS-TO}}

In the absence of scattering, the equation of motion for the reduced
density matrix is

\begin{equation}
i\hbar\,\partial_{t}\rho_{\textbf{k}ss'}=\left[H^{A},\rho\right]_{\textbf{k}ss'}
\end{equation}
We can isolate the perturbation on the right hand side,

\begin{equation}
(i\hbar\,\partial_{t}-\Delta\epsilon_{\textbf{k}ss'})\rho_{\textbf{k}ss'}=\left[V,\rho\right]_{\textbf{k}ss'}
\end{equation}
where $\Delta\epsilon_{\textbf{k}ss'}\equiv\epsilon_{\textbf{k}s}-\epsilon_{\textbf{k}s'}$.

To solve the equation of motion perturbatively, it will be written
in frequency space, order by order in the electric field, using the
expansion of Eq.~\ref{eq:expansion-2}. For every order $n$, the
reduced density matrix will be expressed in terms of its lower order
terms. To alleviate notation and make the recursion relation clearer,
we factorize the electric fields and other common factors by defining,

\begin{align}
\rho_{\textbf{k}ss'}^{(n)}(\omega) & \equiv\int\frac{d\omega_{1}}{2\pi}\dots\frac{d\omega_{n}}{2\pi}\frac{(iq)^{n}E_{\alpha_{1}}(\omega_{1})\dots E_{\alpha_{n}}(\omega_{n})}{\omega_{1}\dots\omega_{n}}\,\nonumber \\
 & \times(2\pi)\,\delta(\omega-\omega_{1}-\dots-\omega_{n})\,\rho_{\textbf{k}ss'}^{\alpha_{1}\dots\alpha_{n}}(\omega_{1},\dots,\omega_{n})\label{defrho}
\end{align}

The goal is now to express the recursion relation between objects
of the form $\rho_{\textbf{k}ss'}^{\alpha_{1}\dots\alpha_{n}}(\omega_{1},\dots,\omega_{n})$.
In the absence of a perturbation, the reduced density matrix is simply
the Fermi-Dirac distribution,

\begin{equation}
\rho_{\textbf{k}ss'}^{(0)}=f(\epsilon_{\textbf{k}s})\,\delta_{ss'}=\frac{\delta_{ss'}}{1+\exp(\frac{\epsilon_{\textbf{k}s}-\mu}{k_{B}T})},
\end{equation}
which, when replaced in Eq.~\ref{J0}, implies $\langle\textbf{J}\rangle^{(0)}=0$,
as expected.

The first order term is 
\begin{equation}
\rho_{\textbf{k}ss'}^{\alpha}(\omega)=\frac{\left[h_{\mathbf{k}}^{\alpha},\rho_{\mathbf{k}}^{(0)}\right]_{ss'}}{\hbar\omega-\Delta\epsilon_{\textbf{k}ss'}},
\end{equation}
and the second order,

\begin{align}
\rho_{\textbf{k}ss'}^{\alpha\beta}(\omega_{1},\omega_{2})= & \frac{1}{\hbar\omega_{1}+\hbar\omega_{2}-\Delta\epsilon_{\textbf{k}ss'}}\nonumber \\
\times & \left(\left[h_{\mathbf{k}}^{\alpha},\rho_{\mathbf{k}}^{\beta}(\omega_{2})\right]_{ss'}+\frac{1}{2}\left[h_{\mathbf{k}}^{\alpha\beta},\rho_{\mathbf{k}}^{(0)}\right]_{ss'}\right)
\end{align}

The pattern is already becoming clear. As an additional example, the
third order term has the form,

\begin{align}
\rho_{\textbf{k}ss'}^{\alpha\beta\gamma}(\omega_{1},\omega_{2},\omega_{3}) & =\frac{1}{\hbar\omega_{1}+\hbar\omega_{2}+\hbar\omega_{3}-\Delta\epsilon_{\textbf{k}ss'}}\times\nonumber \\
 & \left(\left[h_{\mathbf{k}}^{\alpha},\rho_{\mathbf{k}}^{\beta\gamma}(\omega_{2},\omega_{3})\right]_{ss'}+\frac{1}{2}\left[h_{\mathbf{k}}^{\alpha\beta},\rho_{\mathbf{k}}^{\gamma}(\omega_{3})\right]_{ss'}\right.\nonumber \\
 & \left.+\frac{1}{3!}\left[h_{\mathbf{k}}^{\alpha\beta\gamma},\rho_{\mathbf{k}}^{(0)}\right]_{ss'}\right)
\end{align}

Finally, to general order $n$, the perturbative solution to the equation
of motion is recursively expressed as\begin{widetext}

\begin{align}
\rho_{\textbf{k}ss'}^{\alpha_{1}\dots\alpha_{n}}(\omega_{1},\dots,\omega_{n}) & =\frac{1}{\hbar\omega_{1}+\dots+\hbar\omega_{n}-\Delta\epsilon_{\textbf{k}ss'}}\sum_{m=1}^{n}\frac{1}{m!}\left[h_{\mathbf{k}}^{\alpha_{1}\dots\alpha_{m}},\rho_{\mathbf{k}}^{\alpha_{m+1}\dots\alpha_{n}}(\omega_{m+1},\dots,\omega_{n})\right]_{ss'}\label{rho}
\end{align}
\end{widetext}This recursion relation can be unfolded into lengthy
expressions and its structure analyzed in more detail. However, we
shall see that the real value of these expressions lies in their numerical
evaluation, for which a recursion relation is sufficient.

Applying Eq.~\ref{Ref:vexpansion} and Eq.~\ref{defrho} to the
Eq.~\ref{Jn}, the general form of the nonlinear optical response
functions in the velocity gauge is obtained,\begin{widetext}

\begin{align}
\sigma_{\beta\alpha_{1}\dots\alpha_{n}}^{(n)}(\omega_{1},\dots,\omega_{n})=\frac{i^{n}q^{n+1}}{\omega_{1}\dots\omega_{n}}\int\frac{d^{d}\textbf{k}}{(2\pi)^{d}}\sum_{ss'}\sum_{p=0}^{n}\frac{h_{\textbf{k}ss'}^{\beta\alpha_{1}\dots\alpha_{p}}}{p!}\rho_{\textbf{k}s's}^{\alpha_{p+1}\dots\alpha_{n}}(\omega_{p+1},\dots,\omega_{n})\label{responsefunction}
\end{align}
\end{widetext}This form of the nonlinear optical response functions
will still have to undergo the usual symmetrization procedure to ensure
intrinsic permutation symmetry \footnote{It is a consequence of the definition in Eq.~\ref{eq:defresponsefunction}
that only the symmetric part of the response functions contributes
to the integral and is therefore physical.} \cite{shen1984principles}. Albeit trivial, this last step is a bit
cumbersome to write down and will be left implicit.These expressions
are equivalent to the ones we derived in a previous work \cite{ventura2017gauge}
using the length gauge. Although far more complicated, they have their
advantages, which we will discuss in the next section.

The equivalence of the results of the two approaches, length and velocity
gauges, is guaranteed by their being related by a unitary transformation
(see Appendix \ref{B}). It can also be explicitly shown by using
very general sum rules to map the expressions for the nonlinear conductivities
in the velocity gauge to those of the length gauge, order by order,
as shown in an appendix of our previous work \cite{ventura2017gauge}.
The proof for first order responses is nonetheless presented in appendix~A,
as an example of these sum rules, which generalize those derived in
earlier works \cite{sipe1993nonlinear}, no longer rely on commutator
identities (Eq.~\ref{identity}), and are valid for a model with
a finite number of bands.

\section{LENGTH VS VELOCITY GAUGE\label{sec:LENGTH-VS-VELOCITY}}

As usual, there are advantages and disadvantages associated with any
particular choice of gauge. By considering the (exactly equivalent)
forms of the nonlinear conductivities derived in the two gauges, the
strengths and weaknesses of each can be analyzed.

A first look at Eq.~\ref{responsefunction} will immediately bring
out the usual concerns with infrared divergences in the velocity gauge
form, due to all the inverse frequency factors. We emphasize again,
however,  that this expression is equivalent to the one obtained from
the length gauge and therefore these divergences are only apparent.
Manipulating the expressions in the velocity gauge and using a series
of sum rules, it can be shown that Eq.~\ref{responsefunction} is
divergence free. This approach was the one originally pursued \cite{sipe1993nonlinear},
but this use of sum rules became rather pointless after the length
gauge formulation had been developed \cite{aversa1995nonlinear}.
If the sum rules are employed in the velocity gauge to remove apparent
divergences, one will simply arrive at the same expression as obtained
more straightforwardly in the length gauge. This equivalence through
sum rules \cite{ventura2017gauge} does not demand an infinite number
of bands, but it does put a constraint on the use of Eq.~\ref{responsefunction},
namely that the integration must be done over the full FBZ to cancel
divergences. An effective continuum Hamiltonian  describing a portion
of the FBZ\textemdash like the Dirac Hamiltonian for graphene\textemdash ,
 will not suffice, since these sum rules rely on the periodicity in
$\mathbf{k}$ space of the quantities involved.

Having clarified this point, it can still be noted that the velocity
gauge form is considerably more elaborate; less useful not only for
inspection, but in an actual analytical calculation. As an example,
the expression of the second harmonic generation with all components
along the $x$ axis, in the length gauge\cite{ventura2017gauge},
is \footnote{The symbol $\circ$ stands for the Hadamard product: $(A\circ B)_{ss'}=A_{ss'}B_{ss'}$},

\begin{align}
\sigma_{xxx}^{(2)}(\omega,\omega)= & -q^{3}\int\frac{d^{d}\textbf{k}}{(2\pi)^{d}}\sum_{ss'}\frac{h_{\textbf{k}ss'}^{x}}{2\hbar\omega-\Delta\epsilon_{\textbf{k}s's}}\nonumber \\
\times & \left[D_{\mathbf{k}}^{x},\frac{1}{\hbar\omega-\Delta\epsilon}\circ\left[D_{\mathbf{k}}^{x},\rho_{\mathbf{k}}^{(0)}\right]\right]_{s's}\label{SHGE}
\end{align}
while in the velocity gauge,

\begin{align}
\sigma_{xxx}^{(2)}(\omega,\omega) & =-\frac{q^{3}}{\omega^{2}}\int\frac{d^{d}\textbf{k}}{(2\pi)^{d}}\sum_{ss'}\left(h_{\textbf{k}ss'}^{x}\rho_{\textbf{k}s's}^{xx}(\omega,\omega)\frac{}{}\right.\nonumber \\
 & \left.+h_{\textbf{k}ss'}^{xx}\rho_{\textbf{k}s's}^{x}(\omega)+\frac{1}{2}h_{\textbf{k}ss'}^{xxx}\rho_{\textbf{k}s's}^{(0)}\right)\label{SHG}
\end{align}
where we still have to write the density matrix components,

\begin{align}
\rho_{\textbf{k}ss'}^{x}(\omega)= & \frac{\left[h_{\mathbf{k}}^{x},\rho_{\mathbf{k}}^{(0)}\right]_{\textbf{k}ss'}}{\hbar\omega-\Delta\epsilon_{\textbf{k}ss'}}\\
\rho_{\textbf{k}ss'}^{xx}(\omega,\omega)= & \frac{1}{2\hbar\omega-\Delta\epsilon_{\textbf{k}ss'}}\nonumber \\
\times & \left(\left[h_{\mathbf{k}}^{x},\rho_{\mathbf{k}}^{x}(\omega)\right]_{\textbf{k}ss'}+\frac{1}{2}\left[h_{\mathbf{k}}^{xx},\rho_{\mathbf{k}}^{(0)}\right]_{ss'}\right)
\end{align}

This simple example illustrates that there is no advantage in doing
the analytical calculations in the velocity gauge, although inspection
of previous equations shows an interesting point: there are only simple
poles in the velocity gauge form $(\hbar\omega-\Delta\epsilon)^{-1}$,
while in the length gauge, by differentiation, higher order poles
emerge. Still, for analytical calculations, we would advocate the
more transparent and simpler length gauge approach \cite{aversa1995nonlinear,ventura2017gauge}.

The strength of the velocity gauge form lies in the different arrangement
of the commutators. The covariant derivatives are no longer applied
to the density matrix in its recursion relation\footnote{In this aspect, the approach here has similarities with the one employed
in \cite{cheng2014third}.}. Instead, they operate only on the unperturbed Hamiltonian $H_{0}$
in the determination of the functions $h_{\textbf{k}ss'}$ (Eq.~\ref{h-1}),
which are independent of frequency, temperature and chemical potential.

A careful look at the algorithm of the previous section, shows that
for the nonlinear conductivity of order $n$, there are $n+1$ such
functions to compute by successively applying a covariant derivative:
$h_{\textbf{k}ss'}^{\alpha_{1}\dots\alpha_{m}}$ with $m=1,...,n+1$.
In the previous example of second harmonic generation, these would
be $h_{\textbf{k}ss'}^{x}$, $h_{\textbf{k}ss'}^{xx}$ and $h_{\textbf{k}ss'}^{xxx}$.
Further reducing this algorithm to its fundamental ingredients, we
note that these calculations demand only a knowledge of two objects,
which fully define the system under consideration: the dispersion
relation $\epsilon_{\textbf{k}s}$ and the Berry connection $\mathbf{\xi}_{\textbf{k}ss'}$.

Once these $h_{\textbf{k}ss'}^{\alpha_{1}\dots\alpha_{n}}$ functions
are analytically determined, the integrand in Eq.~\ref{responsefunction}
or in Eq.~\ref{SHG} can be numerically evaluated at each $\textbf{k }$value,
independently and quite easily. In fact, the procedure involves evaluating
the analytic $h_{\textbf{k}ss'}^{\alpha_{1}\dots\alpha_{n}}$ functions
and the Fermi-Dirac distribution at the $\textbf{k}$ point and then
computing simple commutators and traces of numeric matrices. It has
no numerical derivatives at all. This is in contrast with the length
gauge, where either the full expression of the response function is
analytically calculated or numerical derivatives have to be applied
in each step of the density matrix recursion relation. Either way,
via the product rule and higher order poles, the number of complicated
terms to evaluate grows very fast with $n$ in the length gauge approach.

For this reason, the form of the nonlinear conductivity in Eq.~\ref{responsefunction},
derived from the velocity gauge, should provide a more efficient numerical
approach. The authors have implemented numerically the expressions
in both gauges and done calculations on the nonlinear conductivity
of monolayer graphene and observed that the computation times were
indeed considerably smaller when Eq.~\ref{responsefunction} was
used. The velocity gauge results are presented in Section~\ref{sec:NUMERICAL-IMPLEMENTATION}.

\section{PHENOMENOLOGICAL RELAXATION PARAMETERS\label{sec:PHENOMENOLOGICAL-RELAXATION-PARA}}

In Eq.~\ref{responsefunction}, like in all previous equations, the
addition of the usual infinitesimal imaginary part to the frequencies,
$\omega\to\omega+i\,0^{+}$, is always implicit, as imposed by causality.
This provide us with well defined relaxation free expressions. From
the numerical standpoint, the imaginary part must always be finite,
but can be taken to be smaller than any other energy scale in the
problem.

However, one is also interested in modeling relaxation processes due
to scattering of electrons with impurities, phonons and other electrons.
A simple phenomenological approach is to add one or more relaxation
parameters to the frequency poles.  A common justification for the
addition of a relaxation parameter $\gamma$ comes from considering
a scattering term in equation of motion\cite{mikhailov2016quantum},

\begin{equation}
i\hbar\,\partial_{t}\rho_{\textbf{k}ss'}=\left[H,\rho\right]_{\textbf{k}ss'}-i\gamma(\rho_{\textbf{k}ss'}-\rho_{\textbf{k}ss'}^{eq})
\end{equation}

If the perturbation is turned off, the density matrix relaxes to the
equilibrium distribution $\rho^{eq}$. In the length gauge approach,
$\rho^{eq}=\rho^{(0)}$. A simple rearrangement\footnote{$V$ here is not the same as in Eq.~\ref{Ref:vexpansion}. It stands
for the perturbation in the length gauge: $V_{\textbf{k}ss'}(t)=-iqE_{\alpha}(t)\,D_{\textbf{k}ss'}^{\alpha}$ }, 

\begin{equation}
(i\hbar\,\partial_{t}+i\gamma-\Delta\epsilon_{\textbf{k}ss'})\rho_{\textbf{k}ss'}=\left[V,\rho\right]_{\textbf{k}ss'}+i\gamma\rho_{\textbf{k}ss'}^{(0)}\label{relaxE}
\end{equation}
makes clear how this will impact the nonlinear conductivities. The
second term on the right hand side of Eq.~\ref{relaxE} is not relevant
to any order $n\neq0$, implying that the only alteration will be
to add an imaginary constant to each pole in frequency space,

\begin{equation}
i\hbar\,\partial_{t}+i\gamma-\Delta\epsilon_{\textbf{k}ss'}\rightarrow\hbar\omega+i\gamma-\Delta\epsilon_{\textbf{k}ss'}
\end{equation}

Again, using the second harmonic generation as an example, the scattering
term will induce the following simple modification of the expression
in Eq.~\ref{SHGE},

\begin{align}
\sigma_{xxx}^{(2)}(\omega,\omega) & =-q^{3}\int\frac{d^{d}\textbf{k}}{(2\pi)^{d}}\sum_{ss'}\frac{h_{\textbf{k}ss'}^{x}}{2\hbar\omega+i\gamma-\Delta\epsilon_{\textbf{k}s's}}\nonumber \\
\times & \left[D_{\mathbf{k}}^{x},\frac{1}{\hbar\omega+i\gamma-\Delta\epsilon}\circ\left[D_{\mathbf{k}}^{x},\rho_{\mathbf{k}}^{(0)}\right]\right]_{s's}
\end{align}

In the case of the velocity gauge, this change is much more complicated,
since the equilibrium distribution depends on the perturbation. Having
in mind the connection between the two gauges, it is easy to see that
the equivalent $\rho^{eq}$ in the velocity gauge equation of motion
should be obtained by an unitary transformation \cite{ventura2017gauge}
of the Fermi-Dirac distribution\footnote{The use of the Fermi-Dirac distribution as $\rho^{eq}$ in the velocity
gauge leads to erroneous results, as already noted in \cite{cheng2017second},
such as the appearance of a nonzero electric current in the absence
of an applied electric field, by means of choosing a constant vector
potential.}. This again translates into an expansion in the vector potential,

\begin{align}
\rho_{\textbf{k}ss'}^{eq}(t) & =\sum_{n=0}^{\infty}\frac{(-q)^{n}\,A_{\alpha_{1}}(t)\dots A_{\alpha_{n}}(t)}{n!\,\hbar^{n}}\nonumber \\
 & \times\left[D_{\mathbf{k}}^{\alpha_{n}},\left[\dots,\left[D_{\mathbf{k}}^{\alpha_{1}},\rho_{\mathbf{k}}^{(0)}\right]\right]...\right]_{ss'}\label{rhoeqA}
\end{align}

If this equilibrium distribution is used, the results obtained will,
indeed, be equivalent to the ones from the length gauge. Nevertheless,
as discussed in the previous section, the main advantage of using
the velocity gauge comes from the absence of derivatives acting on
the density matrix, and once the term in Eq.~\ref{rhoeqA} is added
to the equation of motion, that advantage, and the greater numerical
efficiency of this approach, are lost.

As a consequence, we will follow a different phenomenology, more appropriate
for our computations. To each frequency $\omega_{j}$ we shall add
a constant imaginary part\footnote{This corresponds to choosing $\Gamma^{(n)}=n\gamma$ in the relaxation
parameters in Cheng \emph{et. al.} \cite{cheng2015third}.}: $\hbar\omega_{j}\rightarrow\hbar\omega_{j}+i\gamma$. This method
comes naturally from considering the adiabatic switching of the external
fields. It looks similar to the previous method, but the expression
for second harmonic generation (as would be obtained in the length
gauge)

\begin{align}
\sigma_{xxx}^{(2)}(\omega,\omega) & =-q^{3}\int\frac{d^{d}\textbf{k}}{(2\pi)^{d}}\sum_{ss'}\frac{h_{\textbf{k}ss'}^{x}}{2\hbar\omega+2i\gamma-\Delta\epsilon_{\textbf{k}s's}}\nonumber \\
 & \times\left[D_{\mathbf{k}}^{x},\frac{1}{\hbar\omega+i\gamma-\Delta\epsilon_{\mathbf{k}}}\circ\left[D_{\mathbf{k}}^{x},\rho^{(0)}\right]\right]_{s's}\label{secondgamma}
\end{align}
has now a factor of two in the relaxation parameter that appears in
one of the poles. This might seem like a slight difference, but is
a distinct phenomenology and, as we shall see, even for low values
of $\gamma$ it can lead to completely different results for very
low frequencies $\hbar\omega\ll\gamma$ and in a region of width $\gamma$
around resonances.

\section{NUMERICAL IMPLEMENTATION: THIRD HARMONIC GENERATION IN GRAPHENE\label{sec:NUMERICAL-IMPLEMENTATION}}

In this section, the velocity gauge approach is tested numerically,
by computing the third order conductivity of a material whose nonlinear
optical properties have been the subject of intensive research in
recent years: monolayer graphene \cite{glazov2014high,cheng2014dc,cheng2014third,cheng2015third,cheng2017second,mikhailov2007non,mikhailov2017nonperturbative,marini2017theory,peres2014optical}.

Monolayer graphene is a two dimensional sheet of carbon atoms, displayed
in an honeycomb lattice. It has two atoms per Bravais lattice site.
Here, we shall consider only the simplest nearest neighbor tight binding
model that describes the electronic properties of graphene \cite{neto2009electronic},

\begin{equation}
H_{0}=t\left[\begin{matrix}0 & \phi(\textbf{k})\\
\phi^{*}(\textbf{k}) & 0
\end{matrix}\right]\label{Hmatrix}
\end{equation}
where

\begin{equation}
\phi(\textbf{k})=1+e^{-i\textbf{k}\cdot\textbf{a}_{1}}+e^{-i\textbf{k}\cdot\textbf{a}_{2}}=\left|\phi(\textbf{k})\right|e^{-i\theta(\textbf{k})}
\end{equation}
with $t$ being the hopping parameter and $\textbf{a}_{1}=(1/2,\sqrt{3}/2)\,a$
and $\textbf{a}_{2}=(-1/2,\sqrt{3}/2)\,a$ the basis vectors of the
Bravais lattice. From this model Hamiltonian, the dispersion relation
and the Berry connection can be computed,

\begin{equation}
\epsilon_{\textbf{k}s}=s\,t\left|\phi(\textbf{k})\right|\label{dispersionrelation}
\end{equation}

\begin{equation}
\boldsymbol{\xi}_{\textbf{k}ss'}=-\frac{ss'}{2}\nabla_{\textbf{k}}\theta\label{berryconnection}
\end{equation}
with band index $s=\pm1$ ($-1$ for valence and $+1$ for conduction
band) \footnote{Under the approximation of no overlap between the Wannier orbitals.
Also, an additional constant term in the Berry connection is neglected,
since it is not relevant for frequencies near the Dirac point.}.

The knowledge of the dispersion relation and the Berry connection
is sufficient for a calculation of the nonlinear conductivity, independently
of the gauge. It can be shown that a single independent component
$\sigma_{xxxx}^{(3)}(\omega,\omega,\omega)$ describes third harmonic
generation. We shall confine ourselves to the study of this frequency
component of the third order nonlinear conductivity.

\begin{figure}
\centering \includegraphics[width=1\columnwidth]{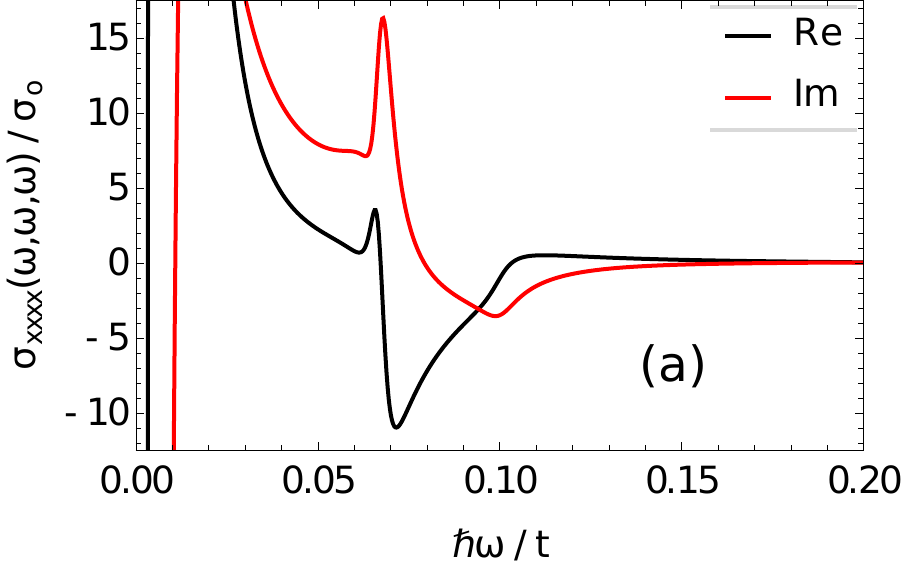} 

\includegraphics[width=1\columnwidth]{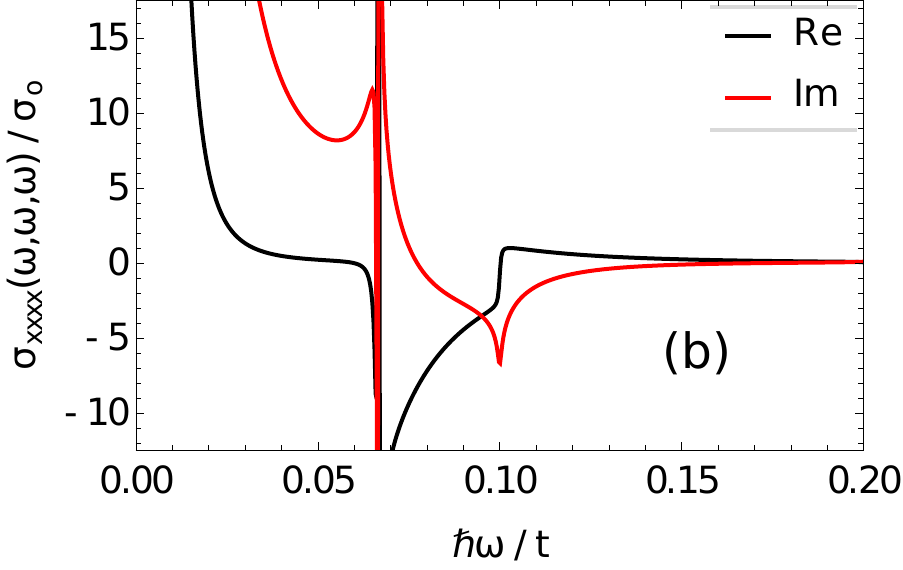} \caption{Frequency dependence of the third order nonlinear conductivity of
graphene, normalized to $\sigma_{0}=3\,q^{4}a^{2}t^{2}/16\pi\hbar\mu^{4}$
(same normalization used in \cite{mikhailov2016quantum}), at zero
temperature. The parameters used were: (a) $\mu/t=0.1$, $\gamma/t=0.011$;
(b) and $\mu/t=0.1$, $\gamma/t=0.001$. The curves were obtained
from a length gauge approach, with the relaxation parameter $\gamma$
introduced via a scattering term in the equation of motion (Eq.~\ref{relaxE}).}
\label{E} 
\end{figure}

\begin{figure}
\centering \includegraphics[width=1\columnwidth]{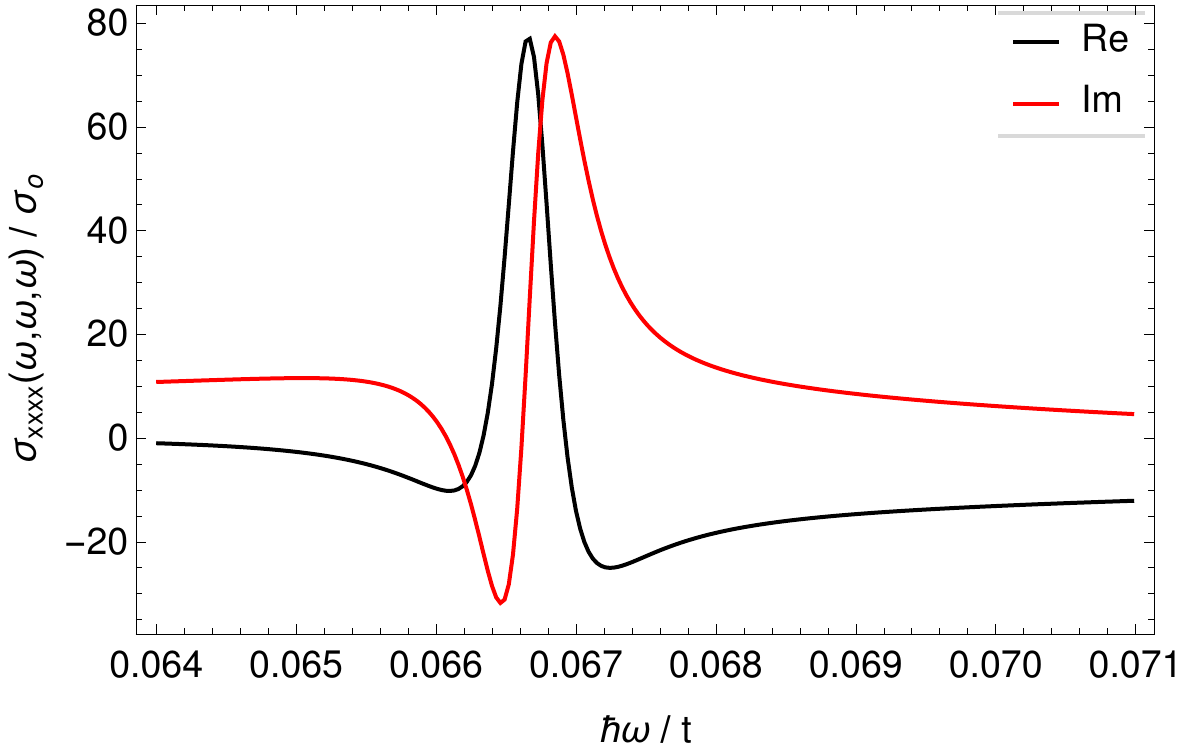} \caption{Anomalous feature of the third order nonlinear conductivity of graphene
from Fig.~\ref{E}(b), in a region near the three photon resonance
$3\hbar\omega=2\mu$.}
\label{E2} 
\end{figure}

Results obtained from the standard length gauge approach are presented
in Fig.~\ref{E}, for frequencies near the Dirac point. In this case,
the analytical form of the third order conductivity in the Dirac point
approximation was calculated and then plotted. Also, $\gamma$ is
introduced via the additional scattering term in the equation of motion
(first type of phenomenological treatment described in the previous
section), as in Mikhailov's work \cite{mikhailov2016quantum}.

These results are in agreement with analogous calculations already
in the literature \cite{cheng2014third,mikhailov2016quantum,cheng2015third},
with the strongest feature present at the three photon resonance $3\hbar\omega=2\mu$.
In particular, Fig.~\ref{E}(a) is identical to Fig. 3(b) and Fig.~5
in refs \cite{cheng2014third} and \cite{mikhailov2016quantum}, respectively.
Fig.~\ref{E}(b) shows a very anomalous behavior at the resonance,
which is always present in a region $\left|3\hbar\omega-2\mu\right|\leq\gamma$
for small but finite $\gamma$. Fig.~\ref{E2} shows a close-up of
this region. This strange feature near resonance is analyzed in detail
in \cite{mikhailov2016quantum}, where it is regarded as a prominent
feature of potential interest, since in practice $\gamma$ is always
finite. However, a more careful analysis shows that despite the unusual
shape of this feature, in the limit $\gamma\rightarrow0^{+}$, the
scattering free result in \cite{cheng2014third} is indeed obtained
(see Appendix~C).

To do the same calculations with the velocity gauge approach developed
here, we evaluate analytically not the full third order conductivity
but only the following functions

\begin{equation}
h_{\textbf{k}ss'}^{x}=\frac{a\,\sin\left(\frac{k_{x}a}{2}\right)C_{ss'}}{\hbar\sqrt{3+2\cos\left(k_{x}a\right)+4\cos\left(\frac{k_{x}a}{2}\right)\cos\left(\frac{\sqrt{3}k_{y}a}{2}\right)}}
\end{equation}

\begin{equation}
h_{\textbf{k}ss'}^{xx}=\frac{a^2\,\cos\left(\frac{k_{x}a}{2}\right)C_{ss'}}{2\hbar^2\sqrt{3+2\cos\left(k_{x}a\right)+4\cos\left(\frac{k_{x}a}{2}\right)\cos\left(\frac{\sqrt{3}k_{y}a}{2}\right)}}
\end{equation}

\begin{equation}
h_{\textbf{k}ss'}^{xxx}=-\frac{a^{2}}{4\hbar^{2}}h_{\textbf{k}ss'}^{x}\qquad h_{\textbf{k}ss'}^{xxxx}=-\frac{a^{2}}{4\hbar^{2}}h_{\textbf{k}ss'}^{xx}
\end{equation}

\begin{align}
C & \equiv t\nonumber \\
\times & \left[\begin{matrix}2\cos\left(\frac{k_{x}a}{2}\right)+\cos\left(\frac{\sqrt{3}k_{y}a}{2}\right) & \mkern-18mu-i\sin\left(\frac{\sqrt{3}k_{y}a}{2}\right)\\
i\sin\left(\frac{\sqrt{3}k_{y}a}{2}\right) & \mkern-18mu-2\cos\left(\frac{k_{x}a}{2}\right)+\cos\left(\frac{\sqrt{3}k_{y}a}{2}\right)
\end{matrix}\right]
\end{align}

All these analytical functions follow from direct evaluation of the
commutators (Eq.~\ref{h-1}) using Eqs.~\ref{dispersionrelation}~and~\ref{berryconnection}.
At this point, the expression in Eq.~\ref{responsefunction} for
the nonlinear conductivity (with $n=3$) can be evaluated by numerical
integration over the full FBZ. The phenomenology adopted here is the
one that follows from adiabatic switching, for the reasons discussed
in the previous section.

\begin{figure}
\centering \includegraphics[width=1\columnwidth]{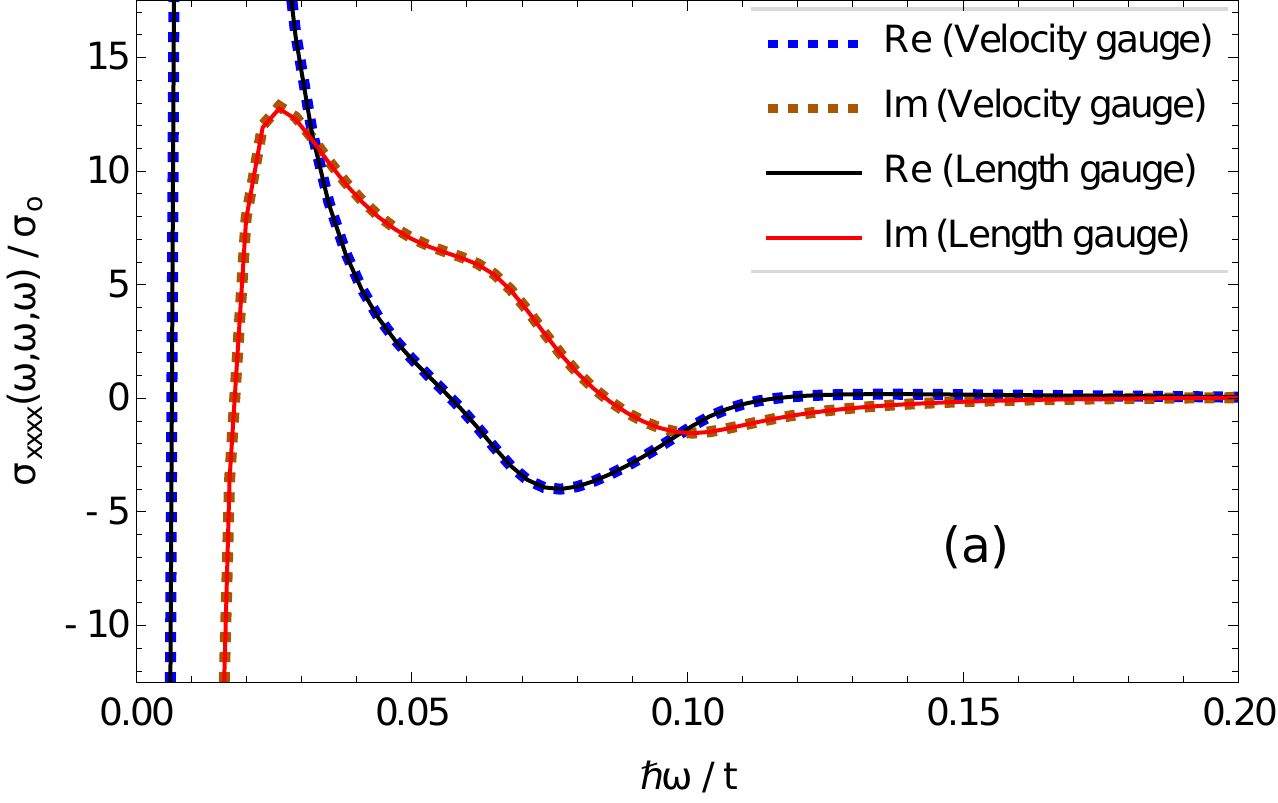}

\includegraphics[width=1\columnwidth]{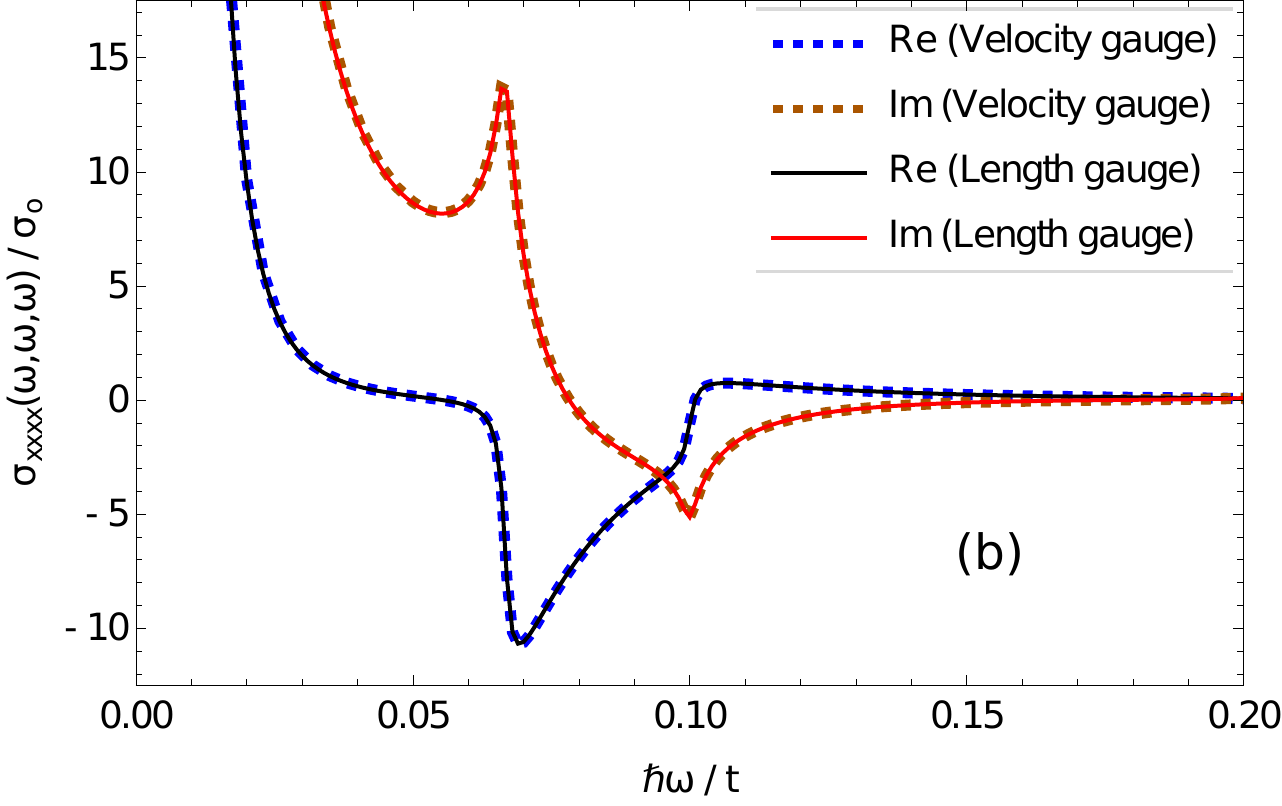} \caption{Frequency dependence of the third order nonlinear conductivity of
graphene, normalized to $\sigma_{0}=3\,q^{4}a^{2}t^{2}/16\pi\hbar\mu^{4}$
(same normalization used in \cite{mikhailov2016quantum}), at zero
temperature. The parameters used were: (a) $\mu/t=0.1$, $\gamma/t=0.011$;
(b) $\mu/t=0.1$, $\gamma/t=0.001$. The solid curves were obtained
from a length gauge approach and the dashed ones from a velocity gauge
calculation. The results are identical. The relaxation parameter $\gamma$
was introduced by adiabatic switching, simply replacing $\hbar\omega\rightarrow\hbar\omega+i\gamma$.}
\label{A} 
\end{figure}

The velocity gauge results are in Fig.~\ref{A}. The curves are markedly
different than the ones obtained from the length gauge, especially
for large $\gamma$. This should came with no surprise, since a different
phenomenological treatment is adopted. To prove that the difference
between the two curves is only due to the way the relaxation parameters
are introduced, the length gauge calculations were performed again,
now with this second type phenomenological treatment (using the third
order analogue of Eq.~\ref{secondgamma}) and included in the plots
of Fig.~\ref{A}. \textit{The results obtained in the two gauges
are identical.}

Of course, as we take the relaxation free limit $\gamma\rightarrow0^{+}$
the results of both phenomenological treatments also coincide. However,
for finite $\gamma$ it may lead to some considerably different behavior
of the nonlinear conductivity. In particular, we note that in the
phenomenology adopted here ($\hbar\omega_{i}\rightarrow\hbar\omega_{i}+i\gamma$),
those anomalous features seen in Fig.~\ref{E} and discussed in \cite{mikhailov2016quantum}
are not present. Instead, a more plausible smooth curve is seen at
the resonances.

As a final remark, we emphasize that the results presented here for
the velocity gauge involved a complete two-band tight binding model
and an integration over the full FBZ, not an effective Dirac Hamiltonian,
as in the case of the length gauge results. This has some consequences.
First, it had been suggested that a possible source for the two order
of magnitude discrepancy between theoretical and experimental results
in graphene could be the use of an effective Hamiltonian \cite{cheng2014third}.
The agreement displayed in Fig.~\ref{A} establishes that the Dirac
point approximation is valid for the range of frequencies adopted
in previous studies\cite{cheng2014third,cheng2015third,mikhailov2016quantum}.
Second, the use of complete bands will allow us to proceed towards
higher frequencies and study spectral regions where the Dirac Hamiltonian
does not give an accurate description.

\section{CONCLUSIONS\label{sec:CONCLUSIONS}}

In summary, a velocity gauge approach to calculations of nonlinear
optical conductivities was developed in this work, within the density
matrix formalism. It was shown that, contrary to common belief, the
results are the same as those from the length gauge, as demanded by
gauge invariance. No difficulties come from applying this velocity
gauge formalism to finite band models.

The velocity gauge provides an efficient algorithm for computing nonlinear
conductivities. The dispersion relation and the Berry connection of
a crystal are the prerequisites and define the crystalline system
under study. A series of commutators (Eq.~\ref{h-1}) can then be
analytically computed. From these, the conductivity can be numerically
evaluated for any frequencies, temperature, chemical potential, relaxation
parameter and to any order, without using any numerical derivatives.

Previously, studies of nonlinear conductivities often involved writing
down analytical expressions for the full conductivity. For third order,
this already becomes rather cumbersome. Analytically, it is only really
tractable for simple effective continuum Hamiltonians (such as the
Dirac Hamiltonian). The velocity gauge approach developed here does
not have such complications, is easily generalizable to higher orders
and is always applied to complete bands. This gauge provide us with
a simple yet powerful method to compute the nonlinear optical response
functions of crystalline systems.

\appendix

\section{Equivalence of linear responses\label{sec:Appendix-A:-Equivalence}}

As an illustration of the general sum rules implicit in a velocity
gauge treatment, the expression for the linear conductivity in the
velocity gauge

\begin{equation}
\sigma_{\alpha\beta}^{(1)}(\omega)=\frac{iq^{2}}{\omega}\int\frac{d^{d}\textbf{k}}{(2\pi)^{d}}\sum_{ss'}\left(\frac{h_{\textbf{k}ss'}^{\beta}\left[h_{\mathbf{k}}^{\alpha},\rho_{\mathbf{k}}^{(0)}\right]_{s's}}{\hbar\omega-\Delta\epsilon_{\textbf{k}s's}}+h_{\textbf{k}ss'}^{\beta\alpha}\rho_{\textbf{k}s's}^{(0)}\right)\label{sigmaA}
\end{equation}
will be shown to be equivalent to the one obtained from a length gauge
approach,

\begin{equation}
\sigma_{\alpha\beta}^{(1)}(\omega)=-iq^{2}\int\frac{d^{d}\textbf{k}}{(2\pi)^{d}}\sum_{ss'}\frac{h_{\textbf{k}ss'}^{\beta}\left[D_{\mathbf{k}}^{\alpha},\rho_{\mathbf{k}}^{(0)}\right]_{s's}}{\hbar\omega-\Delta\epsilon_{\textbf{k}s's}}\label{sigmaE}
\end{equation}
To begin, the Jacobi identity is used to move the covariant derivative
to the density matrix

\begin{align}
\hbar\left[h_{\mathbf{k}}^{\alpha},\rho_{\mathbf{k}}^{(0)}\right] & =\left[\left[D_{\mathbf{k}}^{\alpha},H_{0}\right],\rho_{\mathbf{k}}^{(0)}\right]\nonumber \\
 & =\left[\left[D_{\mathbf{k}}^{\alpha},\rho_{\mathbf{k}}^{(0)}\right],H_{0}\right]+\left[D_{\mathbf{k}}^{\alpha},\left[H_{0},\rho_{\mathbf{k}}^{(0)}\right]\right]\nonumber \\
 & =\left[\left[D_{\mathbf{k}}^{\alpha},\rho_{\mathbf{k}}^{(0)}\right],H_{0}\right]
\end{align}
where in the last step we took into account that the commutator of
two diagonal matrices is zero $\left[H_{0},\rho_{\mathbf{k}}^{(0)}\right]=0$.
This leads to

\begin{equation}
\left[h_{\mathbf{k}}^{\alpha},\rho_{\mathbf{k}}^{(0)}\right]_{ss'}=\hbar^{-1}\left[D_{\mathbf{k}}^{\alpha},\rho_{\mathbf{k}}^{(0)}\right]_{ss'}\Delta\epsilon_{\textbf{k}s's}
\end{equation}
The first term in in parenthesis of Eq.~\ref{sigmaA}, dropping the
\textbf{k} index for simplicity, becomes
\begin{align}
\frac{h_{ss'}^{\beta}\left[h^{\alpha},\rho^{(0)}\right]_{s's}}{\hbar\omega-\Delta\epsilon_{s's}}= & \frac{h_{ss'}^{\beta}\,\hbar^{-1}\left[D^{\alpha},\rho^{(0)}\right]_{s's}\Delta\epsilon_{ss'}}{\hbar\omega-\Delta\epsilon_{s's}}\nonumber \\
= & h_{ss'}^{\beta}\,\hbar^{-1}\left[D^{\alpha},\rho^{(0)}\right]_{s's}\nonumber \\
- & \frac{\omega\,h_{ss'}^{\beta}\left[D^{\alpha},\rho^{(0)}\right]_{s's}}{\hbar\omega-\Delta\epsilon_{s's}}
\end{align}

The second term when replaced in Eq.~\ref{sigmaA} will give the
length gauge result in Eq.~\ref{sigmaE}. The remaining contributions
must therefore be zero and form our sum rule,

\begin{equation}
\frac{iq^{2}}{\omega}\int\frac{d^{d}\textbf{k}}{(2\pi)^{d}}\sum_{ss'}\left(h_{\mathbf{k}ss'}^{\beta}\,\hbar^{-1}\left[D_{\mathbf{k}}^{\alpha},\rho_{\mathbf{k}}^{(0)}\right]_{s's}+h_{\textbf{k}ss'}^{\beta\alpha}\rho_{\textbf{k}s's}^{(0)}\right)=0\label{sumrule}
\end{equation}
This can be further simplified through 
\begin{equation}
h_{ss'}^{\beta\alpha}\equiv\hbar^{-1}\left[D^{\alpha},h^{\beta}\right]_{ss'}
\end{equation}
to

\begin{align}
\sum_{ss'}\left(h_{ss'}^{\beta}\,\hbar^{-1}\left[D^{\alpha},\rho^{(0)}\right]_{s's}+\right. & \left.h_{ss'}^{\beta\alpha}\rho_{s's}^{(0)}\right)\\
= & \hbar^{-1}\sum_{s}\left[D^{\alpha},\,h^{\beta}\rho^{(0)}\right]_{ss}
\end{align}
leading to the form

\begin{equation}
\frac{iq^{2}}{\hbar\omega}\int\frac{d^{d}\textbf{k}}{(2\pi)^{d}}\sum_{s}\left[D_{\mathbf{k}}^{\alpha},h_{\mathbf{k}}^{\beta}\rho_{\mathbf{k}}^{(0)}\right]_{ss}=0\label{sumrule2}
\end{equation}
which can be recognized as a particular case of the sum rules identified
in the appendix A of \cite{ventura2017gauge}. The commutator with
$D^{\alpha}$ can be broken into two pieces, one involving the Berry
connection which is trivially zero (the trace of a proper commutator
is always zero) and another involving a conventional derivative,

\begin{equation}
\frac{iq^{2}}{\hbar\omega}\int\frac{d^{d}\textbf{k}}{(2\pi)^{d}}\sum_{s'}\frac{\partial}{\partial k_{\alpha}}\left(h_{\textbf{k}ss'}^{\beta}\rho_{\textbf{k}s's}^{(0)}\right)=0\label{sumrule3}
\end{equation}
This condition is always true since the functions $h$ and $\rho^{(0)}$
are periodic in reciprocal space. The sum rule (and therefore the
equivalence between the results in the two gauges) is therefore trivially
satisfied as long as the integral is performed over the full FBZ.

\section{Expansion of $H_{0}^{A}$ on the vector potential}

\label{B}

In our previous paper \cite{ventura2017gauge}, we discussed in detail
the equivalence of the length and velocity gauges in the context of
the unperturbed Hamiltonian of Eq.~\ref{free}. In this appendix,
we review this equivalence, using from the start a representation
of the crystal Hamiltonian in terms of a set of bands that can be
finite.

The representation of the position operator in the Bloch basis \cite{blount1962formalisms,ventura2017gauge}requires
the thermodynamic limit. We choose the following normalization for
the Bloch states 
\begin{equation}
\braket{\psi_{\mathbf{k}'s'}}{\psi_{\mathbf{k}s}}=\left(2\pi\right)^{d}\delta\left(\mathbf{k}-\mathbf{k}'\right)\delta_{ss'}
\end{equation}
with the corresponding resolution of the identity 
\begin{equation}
\sum_{s}\int\frac{d^{d}k}{\left(2\pi\right)^{d}}\ket{\psi_{\mathbf{k}s}}\bra{\psi_{\mathbf{k}s}}=\boldsymbol{\hat{1}}
\end{equation}
where the sum over $s$ may be over a finite set of bands. The unperturbed
crystal Hamiltonian is 
\begin{equation}
H_{0}=\sum_{s',s}\int\frac{d^{d}k}{\left(2\pi\right)^{d}}\ket{\psi_{\mathbf{k}s}}\left[H_{0}\right]_{\mathbf{k}ss'}\bra{\psi_{\mathbf{k}s'}}
\end{equation}
with $\left[H_{0}\right]_{\mathbf{k}ss'}=\epsilon_{\mathbf{k}s}\delta_{ss'}$,
and $\epsilon_{\mathbf{k}s}$ the band energies. The perturbation
in the length gauge involves the position operator, 
\begin{equation}
H^{E}=H_{0}-q\mathbf{E}(t)\cdot\mathbf{r}
\end{equation}
Using Blount's result for $\mathbf{r}$ in the thermodynamic limit
\cite{blount1962formalisms}, we showed in our previous paper that
\cite{ventura2017gauge} 
\begin{equation}
\bra{\psi_{\mathbf{k}s}}\mathbf{r}\ket{\psi}=\sum_{s'}i\mathbf{D}_{\mathbf{k}ss'}\braket{\psi_{\mathbf{k}s'}}{\psi}\label{eq:r_operator}
\end{equation}
where the covariant derivative is defined by Eq.~\ref{eq:covDeriv}.

From Eq.~\ref{eq:r_operator}, we can give the position operator
the following representation: 
\begin{equation}
\mathbf{r}=\sum_{s,s'}\int\frac{d^{d}k}{\left(2\pi\right)^{d}}\ket{\psi_{\mathbf{k}s}}i\mathbf{D}_{\mathbf{k}ss'}\bra{\psi_{\mathbf{k}s'}}.
\end{equation}
On first inspection, this equation might suggest that this operator
is diagonal in Bloch momentum; it is not because of the presence of
the gradient with respect to $\mathbf{k}$. Any observable that can
be written as a differential operator acting on the wave function
in a continuous basis will have a similar representation \footnote{The more familiar case is that of the momentum $p_{x}=-i\hbar\int dx\left|x\right\rangle \partial_{x}\left\langle x\right|.$}.

The full single particle Hamiltonian in the length gauge is, therefore,
\begin{equation}
H^{E}=\sum_{s,s'}\int\frac{d^{d}k}{\left(2\pi\right)^{d}}\ket{\psi_{\mathbf{k}s}}\left[\delta_{ss'}\epsilon_{\mathbf{k}s}-iq\mathbf{E}(t)\cdot\mathbf{D}_{\mathbf{k}ss'}\right]\bra{\psi_{\mathbf{k}s'}}
\end{equation}
The velocity gauge is obtained by a time dependent unitary transformation,
\begin{equation}
\ket{\psi_{A}(t)}=S(t)\ket{\psi_{E}(t)}
\end{equation}
\begin{equation}
S(t)=\sum_{s,s'}\int\frac{d^{d}k}{\left(2\pi\right)^{d}}\ket{\psi_{\mathbf{k}s'}}\left[e^{-\frac{q}{\hbar}\mathbf{A}(t)\cdot\mathbf{D}_{\mathbf{k}}}\right]_{s's}\bra{\psi_{\mathbf{k}s}}.
\end{equation}
The time evolution in this gauge is 
\begin{align}
\bra{\psi_{\mathbf{k}s}}i\hbar\frac{d}{dt}\ket{\psi_{A}(t)} & =\bra{\psi_{\mathbf{k}s}}S(t)\hat{H}_{E}S^{\dagger}(t)\ket{\psi_{A}(t)}\nonumber \\
 & +\bra{\psi_{\mathbf{k}s}}i\hbar\frac{dS(t)}{dt}S^{\dagger}(t)\ket{\psi_{A}\left(t\right)}.
\end{align}
The second term is seen to cancel the scalar potential term in the
Hamiltonian, 
\begin{align}
\bra{\psi_{\mathbf{k}s}}i\hbar\frac{dS(t)}{dt}S^{\dagger}(t)\ket{\psi_{A}(t)} & =-i\sum_{s'}q\frac{d\mathbf{A}}{dt}\cdot\mathbf{D}_{\mathbf{k}ss'}\braket{\psi_{\mathbf{k}s'}}{\psi_{A}(t)}\nonumber \\
 & =i\sum_{s'}q\mathbf{E}(t)\cdot\mathbf{D}_{\mathbf{k}ss'}\braket{\psi_{\mathbf{k}s'}}{\psi_{A}(t)}
\end{align}
The velocity gauge Hamiltonian therefore becomes 
\begin{equation}
H^{A}=\sum_{s,s'}\int\frac{d^{d}k}{\left(2\pi\right)^{d}}\ket{\psi_{\mathbf{k}s}}H_{\mathbf{k}ss'}^{A}\bra{\psi_{\mathbf{k}s'}}\label{eq:H_A}
\end{equation}
with 
\[
H_{\mathbf{k}ss'}^{A}\equiv\sum_{r,r'}\left[e^{-\frac{q}{\hbar}\mathbf{A}(t)\cdot\mathbf{D}_{\mathbf{k}}}\right]_{sr}\left[H_{0}\right]_{\mathbf{k}rr'}\left[e^{\frac{q}{\hbar}\mathbf{A}(t)\cdot\mathbf{D}_{\mathbf{k}}}\right]_{r's'}
\]
At this point, we make use of the following identity for any two operators $\hat{B}$
and $\hat{C}$,
\begin{equation}
e^{\hat{C}}\,\hat{B}\,e^{-\hat{C}}=\hat{B}+\left[\hat{C},\hat{B}\right]+\frac{1}{2!}\left[\hat{C},\left[\hat{C},\hat{B}\right]\right]+\dots
\end{equation}

and apply it to Eq.~\ref{eq:H_A} with $\hat{B}=H_{0}$ and $\hat{C}=-\frac{q}{\hbar}\mathbf{A}(t)\cdot\mathbf{D}_{\mathbf{k}}$,
to get 

\begin{align}
H_{\mathbf{k}ss'}^{A}= & \sum_{n=0}^{\infty}\frac{(-q)^{n}\,A_{\alpha_{1}}(t)\dots A_{\alpha_{n}}(t)}{\hbar^{n}\,n!}\nonumber \\
\times & \left[D_{\mathbf{k}}^{\alpha_{n}},\left[\dots,\left[D_{\mathbf{k}}^{\alpha_{1}},H_{0}\right]\right]...\right]_{ss'}\nonumber \\
= & \epsilon_{\mathbf{k}s}\delta_{ss'}\nonumber \\
+ & \sum_{n=1}^{\infty}\frac{(-q)^{n}\,A_{\alpha_{1}}(t)\dots A_{\alpha_{n}}(t)}{\hbar^{n}\,n!}\nonumber \\
\times & \left[D_{\mathbf{k}}^{\alpha_{n}},\left[\dots,\left[D_{\mathbf{k}}^{\alpha_{1}},H_{0}\right]\right]...\right]_{ss'}
\end{align}

\section{Brief note on a phenomenological feature}

\label{C}

In \cite{mikhailov2016quantum}, Mikhailov pointed out that the feature
observed in Fig.~\ref{E2} was due mainly to a term of the form

\begin{equation}
\sim\Im\frac{\gamma}{(3\hbar\omega-2\mu+i\gamma)^{2}}\label{feature}
\end{equation}
In the limit $\gamma\rightarrow0$, this should be considered a distribution,
to be integrated over frequency. One knows that

\begin{equation}
\lim_{\gamma\to0}\Im\frac{1}{3\hbar\omega-2\mu+i\gamma}
\end{equation}
corresponds to a Dirac delta function, so one may ask  to what distribution
corresponds Eq.~\ref{feature}. Since we can write

\begin{align}
\lim\limits _{\gamma\rightarrow0}\Im\frac{\gamma}{(3\hbar\omega-2\mu+i\gamma)^{2}}= & \frac{-1}{3\hbar}\lim\limits _{\gamma\rightarrow0}\Im\partial_{\omega}\frac{\gamma}{3\hbar\omega-2\mu+i\gamma}
\end{align}
upon integration with a test function $f(\omega)$, we obtain a contribution
 $\propto\lim\limits _{\gamma\rightarrow0}\gamma\,f'(2\mu/3\hbar)$
which is always zero. In this limit, the term in Eq.~\ref{feature}
has no weight at all. It will not show up in an integration over $\omega$.
Similar remarks can be made concerning the other resonances in the
nonlinear conductivity, computed via the standard phenomenological
approach; this is how the two phenomenologies discussed in Section~\ref{sec:PHENOMENOLOGICAL-RELAXATION-PARA}
yield different results for any finite $\gamma,$ yet, nevertheless,
become identical in the $\gamma\to0^{+}$ limit. 

\section*{{\normalsize{}{}}}

\bibliographystyle{unsrt}
\bibliography{Nonlinear_Optics}

\begin{thebibliography}{10}

\bibitem{shen1984principles}
Yuen-Ron Shen.
\newblock The principles of nonlinear optics.
\newblock {\em New York, Wiley-Interscience, 1984, 575 p.}, 1984.

\bibitem{sipe1993nonlinear}
JE~Sipe and Ed~Ghahramani.
\newblock Nonlinear optical response of semiconductors in the
  independent-particle approximation.
\newblock {\em Physical Review B}, 48(16):11705, 1993.

\bibitem{aversa1995nonlinear}
Claudio Aversa and JE~Sipe.
\newblock Nonlinear optical susceptibilities of semiconductors: Results with a
  length-gauge analysis.
\newblock {\em Physical Review B}, 52(20):14636, 1995.

\bibitem{ventura2017gauge}
GB~Ventura, DJ~Passos, JMB~Lopes dos Santos, JM~Viana~Parente Lopes, and NMR
  Peres.
\newblock Gauge covariances and nonlinear optical responses.
\newblock {\em Physical Review B}, 96(3):035431, 2017.

\bibitem{genkin1968contribution}
VN~Genkin and PM~Mednis.
\newblock Contribution to the theory of nonlinear effects in crystals with
  account taken of partially filled bands.
\newblock {\em Sov. Phys. JETP}, 27:609, 1968.

\bibitem{hughes1996calculation}
James~LP Hughes and JE~Sipe.
\newblock Calculation of second-order optical response in semiconductors.
\newblock {\em Physical Review B}, 53(16):10751, 1996.

\bibitem{sipe2000second}
JE~Sipe and AI~Shkrebtii.
\newblock Second-order optical response in semiconductors.
\newblock {\em Physical Review B}, 61(8):5337, 2000.

\bibitem{al2014high}
Ibraheem Al-Naib, JE~Sipe, and Marc~M Dignam.
\newblock High harmonic generation in undoped graphene: Interplay of inter-and
  intraband dynamics.
\newblock {\em Physical Review B}, 90(24):245423, 2014.

\bibitem{hipolito2016nonlinear}
F~Hipolito, Thomas~G Pedersen, and Vitor~M Pereira.
\newblock Nonlinear photocurrents in two-dimensional systems based on graphene
  and boron nitride.
\newblock {\em Physical Review B}, 94(4):045434, 2016.

\bibitem{blount1962formalisms}
EI~Blount.
\newblock Formalisms of band theory.
\newblock {\em Solid state physics}, 13:305--373, 1962.

\bibitem{cheng2014dc}
JL~Cheng, N~Vermeulen, and JE~Sipe.
\newblock Dc current induced second order optical nonlinearity in graphene.
\newblock {\em Optics express}, 22(13):15868--15876, 2014.

\bibitem{cheng2015third}
Jin~Luo Cheng, Nathalie Vermeulen, and JE~Sipe.
\newblock Third-order nonlinearity of graphene: Effects of phenomenological
  relaxation and finite temperature.
\newblock {\em Physical Review B}, 91(23):235320, 2015.

\bibitem{nastos2005scissors}
Fred Nastos, Bernd Olejnik, Karlheinz Schwarz, and JE~Sipe.
\newblock Scissors implementation within length-gauge formulations of the
  frequency-dependent nonlinear optical response of semiconductors.
\newblock {\em Physical Review B}, 72(4):045223, 2005.

\bibitem{aversa1994third}
Claudio Aversa, JE~Sipe, M~Sheik-Bahae, and EW~Van~Stryland.
\newblock Third-order optical nonlinearities in semiconductors: The two-band
  model.
\newblock {\em Physical Review B}, 50(24):18073, 1994.

\bibitem{rzkazewski2004equivalence}
K~Rzazewski and Robert~W Boyd.
\newblock Equivalence of interaction hamiltonians in the electric dipole
  approximation.
\newblock {\em Journal of modern optics}, 51(8):1137--1147, 2004.

\bibitem{taghizadeh2017linear}
Alireza Taghizadeh, F~Hipolito, and TG~Pedersen.
\newblock Linear and nonlinear optical response of crystals using length and
  velocity gauges: Effect of basis truncation.
\newblock {\em arXiv preprint arXiv:1710.01300}, 2017.

\bibitem{virk2007semiconductor}
Kuljit~S Virk and JE~Sipe.
\newblock Semiconductor optics in length gauge: A general numerical approach.
\newblock {\em Physical Review B}, 76(3):035213, 2007.

\bibitem{cheng2014third}
JL~Cheng, Nathalie Vermeulen, and JE~Sipe.
\newblock Third order optical nonlinearity of graphene.
\newblock {\em New Journal of Physics}, 16(5):053014, 2014.

\bibitem{berry1984quantal}
Michael~V Berry.
\newblock Quantal phase factors accompanying adiabatic changes.
\newblock In {\em Proceedings of the Royal Society of London A: Mathematical,
  Physical and Engineering Sciences}, volume 392, pages 45--57. The Royal
  Society, 1984.

\bibitem{xiao2010berry}
Di~Xiao, Ming-Che Chang, and Qian Niu.
\newblock Berry phase effects on electronic properties.
\newblock {\em Reviews of modern physics}, 82(3):1959, 2010.

\bibitem{mikhailov2016quantum}
Sergey~A Mikhailov.
\newblock Quantum theory of the third-order nonlinear electrodynamic effects of
  graphene.
\newblock {\em Physical Review B}, 93(8):085403, 2016.

\bibitem{cheng2017second}
JL~Cheng, N~Vermeulen, and JE~Sipe.
\newblock Second order optical nonlinearity of graphene due to electric
  quadrupole and magnetic dipole effects.
\newblock {\em Scientific Reports}, 7, 2017.

\bibitem{glazov2014high}
MM~Glazov and SD~Ganichev.
\newblock High frequency electric field induced nonlinear effects in graphene.
\newblock {\em Physics Reports}, 535(3):101--138, 2014.

\bibitem{mikhailov2007non}
SA~Mikhailov.
\newblock Non-linear electromagnetic response of graphene.
\newblock {\em EPL (Europhysics Letters)}, 79(2):27002, 2007.

\bibitem{mikhailov2017nonperturbative}
SA~Mikhailov.
\newblock Nonperturbative quasiclassical theory of the nonlinear electrodynamic
  response of graphene.
\newblock {\em Physical Review B}, 95(8):085432, 2017.

\bibitem{marini2017theory}
A~Marini, JD~Cox, and FJ~Garc{\'\i}a de~Abajo.
\newblock Theory of graphene saturable absorption.
\newblock {\em Physical Review B}, 95(12):125408, 2017.

\bibitem{peres2014optical}
NMR Peres, Yu~V Bludov, Jaime~E Santos, Antti-Pekka Jauho, and MI~Vasilevskiy.
\newblock Optical bistability of graphene in the terahertz range.
\newblock {\em Physical Review B}, 90(12):125425, 2014.

\bibitem{neto2009electronic}
AH~Castro Neto, F~Guinea, Nuno~MR Peres, Kostya~S Novoselov, and Andre~K Geim.
\newblock The electronic properties of graphene.
\newblock {\em Reviews of modern physics}, 81(1):109, 2009.

\end{thebibliography}

\end{document}